\documentclass[sigconf,nonacm,screen]{acmart}
\setcopyright{none}
\renewcommand\footnotetextcopyrightpermission[1]{}
\usepackage{subcaption}

\begin{document}

\title{Efficient Estimation of Shortest-Path Distance Distributions to Samples in Graphs}

\author{Alan Zhu}
\email{aczhu@berkeley.edu}
\affiliation{%
  \institution{University of California, Berkeley}
  \city{Berkeley}
  \state{California}
  \country{USA}
}

\author{Jiaqi Ma}
\email{jiaqima@illinois.edu}
\affiliation{%
  \institution{University of Illinois Urbana-Champaign}
  \city{Urbana}
  \state{Illinois}
  \country{USA}}

\author{Qiaozhu Mei}
\email{qmei@umich.edu}
\affiliation{%
  \institution{University of Michigan}
  \city{Ann Arbor}
  \state{Michigan}
  \country{USA}}

\begin{abstract}
As large graph datasets become increasingly common across many fields, sampling is often needed to reduce the graphs into manageable sizes. This procedure raises critical questions about representativeness as no sample can capture the properties of the original graph perfectly, and different parts of the graph are not evenly affected by the loss. Recent work has shown that the distances from the non-sampled nodes to the sampled nodes can be a quantitative indicator of bias and fairness in graph machine learning. However, to our knowledge, there is no method for evaluating how a sampling method affects the distribution of shortest-path distances without actually performing the sampling and shortest-path calculation.

In this paper, we present an accurate and efficient framework for estimating the distribution of shortest-path distances to the sample, applicable to a wide range of sampling methods and graph structures. Our framework is faster than empirical methods and only requires the specification of degree distributions. We also extend our framework to handle graphs with community structures. While this introduces a decrease in accuracy, we demonstrate that our framework remains highly accurate on downstream comparison-based tasks. Code is publicly available at \url{https://github.com/az1326/shortest_paths}.
\end{abstract}

\maketitle

\balance

\section{Introduction}
\label{sec:intro}

Graph sampling is commonly used in a variety of fields that are concerned with large scale graph data such as the Web graph or real-world social networks. In many cases, accessing or processing the full graph is impractical or impossible, and one has to effectively reduce the size of the graph (often through sampling the nodes) so that sophisticated data science methods can be practically deployed. For example, Web crawlers travel the graph of Web pages to model the Internet~\cite{boldi2004your}, epidemiologists sample a population to model the spread of diseases~\cite{cohen2003efficient}, and biologists sample cell interactions to understand behavior of cellular networks~\cite{aittokallio2006graph}. Sampling can also be used to simplify complex graph structures that would otherwise be difficult to work with~\cite{kurant2012coarse}. In particular, when dealing with large social networks, sampling reduces the size and structural complexity of the graph and allows for easier analysis~\cite{wang2011understanding}. 

However, graph sampling is not limited to selecting nodes at random, even though this approach may be appropriate in some cases. Ideally, the sampling process should be tailored to the specific research question while minimizing any unwanted bias. For example, sampling hard-to-reach communities poses unique challenges in that those who are successfully sampled may not be representative of the larger population~\cite{salganik2004sampling}. In addition, samples would ideally retain relevant properties of the original graph, but retention can be difficult~\cite{stumpf2005subnets}. Poor sampling practices can introduce unwanted bias~\cite{maiya2011benefits}, especially in more complex graphs where the importance of nodes is not necessarily equal~\cite{stutzbach2006sampling}. Carefully choosing a sampling method can help address these representation concerns.

A crucial way to understand the representativeness of graph sampling methods is through the perspective of the distribution of shortest-path distances (DSPD) from other nodes to a sample.
The DSPD provides valuable insights into how well the sample covers the graph, a measure of representativeness known as network reach~\cite{maiya2011benefits}.
Additionally, in graph neural networks~\cite{scarselli2008graph}, distance from sample has been shown to be an indicator of fairness~\cite{ma2021subgroup}, as machine learning models perform better on nodes closer to the labeled nodes than on those further away.
When dealing with hard-to-reach communities, models perform worse on those nodes further away from the sample~\cite{heckathorn2017network}. However, determining the DSPD can be difficult: the straightforward way to perform sampling then empirically calculate the shortest-path distances can be computationally expensive and requires full knowledge of the graph structure.

This paper provides, to our knowledge, the first framework for efficiently and accurately estimating the DSPD to a sample without accessing the full graph structure. Such a framework will help inform graph sampling decisions and enable future research into the DSPD to a sample. We demonstrate that our framework is more efficient compared to naive methods even assuming full knowledge of the graph, and that it is highly accurate for graphs without community structures. While we observe reduced accuracy for graphs with community structures, we show that that our framework remains highly reliable for downstream comparison tasks---a common use case when using our framework to inform sampling decisions.

The rest of the paper is organized as follows: Section \ref{sec:related} surveys existing works related to graph sampling and shortest-path distances. Section \ref{sec:framework} describes our framework for estimating the DSPD. Section \ref{sec:exp} describes the experiments we run to verify the accuracy and efficiency of our framework. Section \ref{sec:results} presents the results of our experiments. Section \ref{sec:conclusion} concludes the paper.

\section{Related Work}
\label{sec:related}

\subsection{Graph Sampling}

Graph sampling is a common technique used in cases where working with a full graph is impractical or impossible. These cases can arise when collecting data from the full network or working with the full network is computationally infeasible, or in social science contexts when working with a hard-to-reach communities, to name a few.
\citet{maiya2011benefits} discusses how well a sample covers a graph through \textit{network reach}, measuring it through \textit{discovery quotient}, which is the proportion of nodes that are in the sample or is a neighbor of a sample. However, we examine a more general idea via the full distribution of distances.

The effect of sampling methods on various metrics is also well-studied.
\citet{hu2013survey} conduct a survey of graph sampling techniques and provide ad hoc analysis of the impacts of the sampling techniques.
\citet{costenbader2003stability} look at the stability of centrality metrics under sampling. \citet{stumpf2005subnets} conclude that the scale-free property of a graph would not be preserved under random sampling.
These studies all look at whether certain properties of the original graph would be preserved by the sampled graph. This paper instead examines the representativeness of nodes in a sample using the DSPD of other nodes to that sample. This property is not intrinsic to the original graph, but depends on both the graph and the sample, and therefore is not something to be preserved ``before and after'' sampling.

\subsection{Shortest-Path Distance}
Classical network science literature measures the mean of shortest-path distances in a graph and uses it to categorize networks \cite{watts1998collective}. 
A few studies look at the DSPD beyond the mean, including \citet{katzav2018distribution}, 
\citet{ventrella2018modeling}, and \citet{katzav2015analytical}. \citet{katzav2018distribution} obtain an analytical expression for the DSPD between nodes in subcritical Erd\"os-R\'enyi graphs, while \citet{katzav2015analytical} do so for Erd\"os-R\'enyi graphs in general. \citet{ventrella2018modeling} propose models that can be used to find the DSPD for scale-free networks. These studies all model the DSPD in the full graph, while we do so for the DSPD \textit{to a sample of nodes} selected from the full graph.

Recent work has made note of the importance of the DSPD to a sample. \citet{ma2021subgroup} find that the distance of an unlabeled node to the subgroup of labeled nodes in a graph is a good indicator of the performance of a graph neural network on that node, which helps motivate our investigation of the DSPD.

\section{Theoretical Framework}
\label{sec:framework}

In this section, we describe our framework for efficiently estimating the DSPD. Our framework works primarily within the configuration model~\citep{newman2003structure} setting, where the degree distribution of the nodes in the graph is specified. Notably, the configuration model does not account for community structure as it assumes the degree of each node is drawn independently of all the others from the specified distribution. We thus also present a method to account for community structure using our framework within the Stochastic Block Model (SBM)~\cite{holland1983stochastic} setting.

\subsection{Preliminaries}
\label{sec:preliminaries}

As preliminaries, we begin with a simpler question: the DSPD to a \textit{single random node} in a graph. This is equivalent to determining the DSPD in the entire graph. \citet{nitzan2016distance} derive such a distribution analytically in the configuration model setting.

Intuitively, the distribution is obtained by looking at ``shells'' around a single node recursively. The probability that the shortest path distance between two random points $i$ and $j$ is greater than $\ell$ can be calculated recursively with respect to $\ell$ as the probability that shortest-path distance between every neighbor of $i$ and $j$ is greater than $\ell-1$. This can be determined from the probability that the distance between any two nodes is greater than $\ell-1$ and the distribution of the number of neighbors of $i$. An additional complication is that the degree distribution of a randomly selected node is different from the degree distribution of a node known to have a neighbor, and this is resolved by maintaining two recursive equations.

Begin with the chain rule: for graphs with $N$ nodes,
\[P(d > \ell) = P(d > 0) \prod_{\ell'=1}^\ell m_{N, \ell'}, \]
where $d$ is the shortest-path distance between a pair of nodes $i, j$ and $m_{N, \ell} =  P(d > \ell | d > \ell-1)$ in a graph with $N$ nodes.

First,
\[m_{n, \ell} = \sum_{k=1}^{n-2} p(k) (\tilde{m}_{n-1,\ell-1})^k.\]
Here $p(k)$ is the degree distribution. For a pair of nodes $i,j$, let $r$ be a neighbor of $i$. $\tilde{m}_{n-1,\ell-1}$ is the probability that in a graph of size $n-1$ (i.e., the graph on the LHS but excluding $i$), the shortest-path distance from $r$ to $j$ is greater than $\ell-1$ given that the distance is greater than $\ell - 2$.

Intuitively, we are given that the distance from $i$ to $j$ is greater than $\ell-1$, so we know the distance from every neighbor of $i$ to $j$ is greater than $\ell-2$. To find the probability that the distance from $i$ to $j$ is greater than $\ell$ we thus additionally require that the distance from every neighbor of $i$ to $j$ to be greater than $\ell-1$.

Next,
\[\tilde{m}_{n, \ell} = \sum_{k=1}^{n-2} \frac{k}{c} p(k) (\tilde{m}_{n-1, \ell-1})^{k-1},\]
where $c = \sum_{k=1}^\infty kp(k)$ is a normalizing constant. Here the distribution $\frac{k}{c} p(k)$ is used as the degree distribution of $i$'s neighbor $r$ is not drawn from $p(k)$: the probability of a node being a neighbor of $i$ is proportional to its degree. The exponent is $k-1$ as one of $r$'s neighbors is $i$.

The base cases are
\[m_{n,1} = \sum_{k=1}^{n-1} p(k) \left(1 - \frac{1}{n-1}\right)^k,\]
and
\[\tilde{m}_{n,1} = \sum_{k=1}^{n-1} \frac{k}{c}p(k) \left(1 - \frac{1}{n-1}\right)^{k-1}.\]

We may now recursively solve for $m_{N,\ell'}$ for $\ell' = 1,2,\cdots,\ell$ and calculate $P(d > \ell)$.

\subsection{Multi-Node Sample Framework}

Our proposed framework extends the analysis above to the DSPD to a \emph{collection} of sampled nodes, rather than a single random node. For a sample size of $s$, the problem is more complex than simply considering the distribution of the minimum among $s$ samples from the single node DSPD, as the distances to each sample node are not necessarily independent. Additionally, our framework must accommodate cases where sample nodes are not selected randomly.

To overcome these challenges, we retain the idea of looking at shells centered around a single node, but view the center of the shell as a \textit{supernode} consisting of all the nodes in the sample contracted into one. More formally, let a graph be $G = (V, E)$, where $V$ is the set of nodes and $E\subseteq V\times V$ is the set of edges. Denote a set of sample nodes as $S\subseteq V$, all the edges of $G$ involving a node in $S$ as $E_S \subseteq E$. Let $\mathcal{N} := \{v\in V\setminus S \mid (u, v) \in E\}$, which is the set of neighbors of nodes in $S$ that are outside $S$. Then the graph (which we call a \textit{contracted graph}) after contracting the sample $S$ into a supernode $u_S$ is $G' = (V', E')$, where 
\[V' := V \setminus S \cup \{u_S\}, E' := E \setminus E_S \cup \{(u_S, v) \mid v\in N\}).\]

We then need to adjust the formula for $m_{n,\ell}$. In particular, instead of the degree distribution $p(k)$ in the original graph $G$, we draw from the degree distribution of the supernode $u_S$ in the contracted graph $G'$ (i.e., the distribution of the degree of a supernode in a contracted graph), denoted as $p_S(k)$. Additionally, let $N'$ be the number of nodes in the contracted graph. Then,
\[
P(d > \ell) = P(d > 0) \prod_{\ell'=1}^\ell m_{N', \ell'},
\]
as before. However, the recursions are now
\[
    m_{n, \ell} = \sum_{k=1}^{n-2} p_S(k) (\tilde{m}_{n-1,\ell-1})^k,
\]
and
\[
    \tilde{m}_{n, \ell} = \sum_{k=1}^{n-2} \frac{k}{c} p(k) (\tilde{m}_{n-1, \ell-1})^{k-1}.
\]
The base cases are
\[
    m_{n,1} = \sum_{k=1}^{n-1} p_S(k) \left(1 - \frac{1}{n-1}\right)^k,
\]
and
\[
    \tilde{m}_{n,1} = \sum_{k=1}^{n-1} \frac{k}{c}p(k) \left(1 - \frac{1}{n-1}\right)^{k-1}.
\]

It then remains to determine $p_S$, which depends on the structure of $G$ and the sampling method used to draw $S$. Examples of such determinations for the graphs and sampling methods involved in our experiments are presented in Section \ref{sec:sampling_methods}. Finally we may solve the recursion to obtain the DSPD to the collection of sampled nodes.

\subsection{Community Structure}

Note that our framework as presented above is not able to account for community structures: it assumes that the degree of a node is independent of all other nodes in the graph. We now present an extension to our framework that is able to account of community structure under the SBM setting. Here, graphs are split into blocks, with each edge within a block appearing with probability $p_1$ and each edge across two blocks appearing with probability $p_2$. Each block then models a community, and with $p_1 > p_2$ the connections are denser within a community than across communities. We leave applications on other community structure models as future work.

Challenges thus arise as there are two degree distributions to track: the degree within a block and the degree across a block. This is important as in our recursion we examine the degree distribution of a neighbor node, and now that distribution is different depending on whether that neighbor relation is via an within-block or across-block edge. To resolve this we keep an additional set of recursions. For an SBM graph $G$, contracted to $G'$ with $N'$, denote $p_w$ as the within-block degree distribution (i.e., degree distribution counting only within-block edges) and $p_a$ as the across-block degree distribution.

As before,
\[
P(d > \ell) = P(d > 0) \prod_{\ell'=1}^\ell m_{N', \ell'}.
\]
However, now
\begin{align*}
    m_{n, \ell} = \sum_{k_w + k_a < n-1} &p_{Sw}(k_w) p_{Sa}(k_a) \\
    &\left(\tilde{m}_{n-1, \ell-1}^{w}\right)^{k_w} \left(\tilde{m}_{n-1, \ell-1}^{a}\right)^{k_a},
\end{align*}
where $p_{Sw}$ is the sample supernode degree distribution counting only within-block edges in $G'$, $p_{Sa}$ is the sample supernode degree distribution counting only outside-block edges in $G'$, $\tilde{m}_{n-1, \ell-1}^w$ is the probability that a node in the next shell reached via a within-block edge is greater than $\ell-1$ away from the node given that it is greater than $\ell - 2$ away, and $\tilde{m}_{n-1, \ell-1}^a$ is the probability that a node in the next shell reached via an across-block edge is greater than $\ell-1$ away from the node given that it is greater than $\ell - 2$ away. The recursions for $\tilde{m}^w$ and $\tilde{m}^a$ are
\begin{align*}
    \tilde{m}_{n, \ell}^{w} = \sum_{k_w + k_a < n-1} &\frac{k_w p_w(k_w)}{c_w} p_a(k_a) \\
    &\left(\tilde{m}_{n-1, \ell-1}^{w}\right)^{k_w-1} \left(\tilde{m}_{n-1, \ell-1}^{a}\right)^{k_a}, \\
    \tilde{m}_{n, \ell}^{o} = \sum_{k_w + k_a < n-1} &p_w(k_w) \frac{k_a p_a(k_a)}{c_a} \\
    &\left(\tilde{m}_{n-1, \ell-1}^{w}\right)^{k_w} \left(\tilde{m}_{n-1, \ell-1}^{a}\right)^{k_a-1},
\end{align*}
with $c_w$ and $c_a$ being $\sum_{0 < k < n-1}kp_w(k)$ and $\sum_{0 < k < n-1}kp_a(k)$, the normalizing constants after weighting $p_w$ and $p_a$ by the degree. Depending on whether the node in the next shell was arrived at via a within-block edge or across-block edge, the within-block or across-block degree distribution is the weighted weighted version rather than the unweighted degree distribution, respectively. The base cases are then
\begin{align*}
    m_{n, 1} &= \sum_{k_w + k_a < n} p_{Sw}(k_w) p_{Sa}(k_a) \left(1 - \frac{1}{Nn-1}\right)^{k_w + k_a}, \\
    \tilde{m}_{n,1}^{w} &= \sum_{k_w + k_a < n} \frac{k_w p_w(k_w)}{c_w} p_a(k_a) \left(1 - \frac{1}{n-1}\right)^{k_w + k_a - 1}, \\
    \tilde{m}_{n,1}^{a} &= \sum_{k_w + k_a < N} p_w(k_w) \frac{k_a p_a(k_a)}{c_a} \left(1 - \frac{1}{n-1}\right)^{k_w + k_a - 1}.
\end{align*}

Similarly, it remains to determine $p_{Sw}$ and $p_{Sa}$, which depends on $p_1, p_2$ and the sampling method used to draw $S$. Details are presented in Section \ref{sec:sampling_methods}. Finally we may solve the recursion to get the results.\\

Notably, these recursions only require the degree distribution of the original graph and the distribution of the degree of the sample supernode in the contracted graph. Such distributions can be derived from assumptions on the generative model of the graph and the sampling method, without access to the full graph structure or sample necessary for empirically calculating the DSPD.

\section{Experiments}
\label{sec:exp}

To verify the accuracy and efficiency of our framework, we run extensive experiments on a variety of settings, including different graph configurations, graph sizes, sampling methods, and sample sizes. We additionally evaluate the usefulness of our framework on the downstream task of comparing mean distances. We focus our experiments on synthetic graphs as our framework works using the specification of the degree distribution. We thus focus on accurately estimating the DSPD assuming a correct model; we view the task of deriving an accurate model of a real-world graph as a separate task. Our experiments are implemented in NetworkX~\cite{SciPyProceedings_11} and NumPy~\cite{harris2020array}.

\subsection{Graph Configurations}

We focus our experiments on three types of graphs: binomial graphs, power-law graphs~\cite{barabasi1999emergence}, and SBM graphs. The first two graphs do not model community structure, while SBM graphs do.

\subsubsection{Binomial Graphs}

Binomial graphs are graphs with $N$ nodes and probability parameter $p$, where every possible edge is in the graph with probability $p$, independent of every other edge. The degree distribution is given by $p(k) = \binom{N-1}{k}p^k(1-p)^{N-1-k}$, the $\text{Binomial}(N-1,p)$ distribution. Binomial graphs are simple and intuitive to implement, making them attractive to analyze, but are not as flexible as more complex models.

We run experiments on settings of $(N,p) = (20000, 0.0005)$, $(40000, 0.00025)$, and $(100000, 0.0001)$, wherein the graphs get larger but the expected degree of each node remains constant.

\subsubsection{Power-Law Graphs}

The power-law model is a way to generate scale-free graphs, where the degree distribution follows a power-law distribution. For a graph with $N$ nodes and power-law parameter $\gamma$, the degree distribution is $p(k) = ak^{-\gamma}$ where $a$ is a normalizing constant. Minimums and maximums can be enforced on the support of the distribution. Many social networks are scale-free, making power-law graphs a potentially good way to replicate real-world graphs.

\citet{barabasi1999emergence} suggest that values of $\gamma$ in large real-world graphs commonly fall between $2$ and $3$; we thus consider two configurations: Power Law A, with $\gamma = 2$ and support $6 \leq k \leq 29$; and Power Law B, with $\gamma = 3$ and support $6 \leq k \leq 19$. The supports were chosen empirically to allow for opposite preferences for sampling method when considering the downstream task, allowing us to better evaluate our framework's performance on downstream tasks. For each configuration of power law graph, we run experiments with $N = 20000, 40000,$ and $100000$.

\subsubsection{Stochastic Block Model}

The SBM generates graphs resembling real-world social networks with community structures. The graph is split into blocks with each edge within a block appearing with probability $p_1$ and each edge across two blocks appearing with probability $p_2$. In practice, blocks can be of different sizes, but for simplicity we assume each block is of the same size; this allows the degree distributions of each node to be the same, which is necessary for our framework. Then, for a graph with $B$ blocks and $N_B$ nodes per block, the within-block degree distribution is $p_w(k)=\binom{N_B-1}{k}p_1^{k}(1-p_1)^{N_b-1-k}$ and the across-block degree distribution is $p_a(k) = \binom{(B-1)N_B}{k}p_2^k (1-p_2)^{(B-1)N_B-k}$.

To inform our design decision for the SBM graphs, we examine the Facebook Large Page-Page Network graph~\cite{rozemberczki2019multiscale} from the Stanford Network Analysis Project (SNAP)~\cite{snapnets}, a graph with 22,470 nodes representing official pages on Facebook, with 171,002 undirected edges representing mutual likes between pages. We run Clauset-Newman-Moore greedy modularity maximization~\cite{clauset2004finding} to partition the graph into $130$ communities, each with an average block size of about $172$. We next determine the within-block and across-block density to match the number of edges after enforcing equal block sizes.

We obtain graphs of different sizes by scaling the number of blocks,  generating SBM graphs with $p_1=0.08323$, $p_2=0.00004718$, $N_B = 172$, and $B = 130, 260,$ and $650$. These settings correspond approximately to graphs of size $N = 22360, 44720,$ and $111800$, approximately matching the size settings of other graph configurations. When presenting results on these graphs, we will refer to these as graph size $20000, 40000,$ and $100000$ for consistency between graph configurations.

\subsection{Sampling Methods}
\label{sec:sampling_methods}

We experiment using two common sampling methods: random sampling and snowball sampling. Three sample sizes are used: $s=200, 400$ and $1000$. These sample sizes roughly correspond to $1\%$ of the nodes for each size of graph. Applications and evaluations on other sampling methods are left as future work.

\subsubsection{Random Sampling}

Random sampling is a straightforward way to sample from a graph. To generate a sample of size $s$, we select $s$ nodes from the graph uniformly at random, without replacement.

The sample supernode degree distribution $p_S$ can then be determined easily.
Let $K_i$ be the random variable representing the degree of the $i$th node in the sample with pmf $p(k)$.
For a sample of size $s$, we are then interested in the distribution of $K_S = \sum_{i=1}^s K_i$, or the sum of $s$ independent draws of $K_i$. This distribution can be efficiently calculated using polynomial exponentiation: let the polynomial $q(x) = \sum_{k=0}^\infty p(k) x^k$. 
In practice, as $K_i$ has finite support, $q(x)$ is a finite polynomial. Then, $q_S(x) = q(x)^s$, and $p_S(k)$ is simply the coefficient on $x^k$ in $q_S(x)$.
In the case of SBM graphs, the calculations are similar, except that the process is performed twice: once to determine $p_{Sw}(k)$ using $p_w(k)$ and once to determine $p_{Sa}(k)$ using $p_a(k)$.

A key assumption in these derivations is that there are negligible edges between two nodes in the sample, i.e. that every edge out of a node in the sample remains an edge counted in the degree of the contracted supernode. In cases where the sample size is small and the graph is large, this will generally be true.

\subsubsection{Snowball Sampling}

Snowball sampling~\cite{goodman1961snowball} is a form of sampling commonly used in sociology when sampling on hard-to-reach communities~\cite{hu2013survey}, where random sampling is not feasible.
In snowball sampling, we begin with a seed sample, then produce a frontier of all nodes adjacent to the seed. At each step, a node in the frontier is removed from the frontier and added to the sample with a certain probability. If the node is added, it is removed from the frontier and the frontier is expanded with that node's neighbors. This simulates the process of using referrals to find new subjects, which may be necessary if the population being sampled is stigmatized and standard surveys are ineffective. In our experiments, we begin with a single random seed node, nodes are added to the sample with probability $0.5$, and if the frontier is exhausted without having reached the sample size $s$, the process is continued from a new random seed node.

The process of deriving the sample supernode degree distribution $p_S$ becomes more complicated.
Under snowball sampling, all nodes except the seed nodes are reached via exploring an edge. Thus, the probability that a node is in the snowball sample is proportional to its degree. Moreover, each node in the sample contributes $2$ fewer degrees to the supernode than its actual degree as the edge used to reach it is fully within the supernode. We can approximate the degree distribution of the supernode as follows: let the degree distribution be $p(k)$ and let $p'(k) = \frac{k}{c}p(k)$ where $c$ is a normalizing constant such that $p'(k)$ is the degree distribution of a node reached via exploring an edge. For a sample of size $s$, we may now proceed as in the random sampling case: let the polynomial $q(x) = \sum_{k=0}^\infty p'(k)x^k$. Then $q_S(x) = q(x)^s$, and $p_S(k)$ is approximately the coefficient on $x^{k+2s}$. The correction of $2s$ accounts for degree contribution of $2$ less from each node in the sample. Note that this is an approximation as it assumes every node is reached via snowball sampling, while some nodes are indeed seed nodes. However, when the sample size is large and the retention probability is high relative to the degree distribution, this approximation is accurate.

\paragraph{SBM Graphs} For SBM graphs, the process is more complicated still as we must track whether the sample node is reached via a within-block or across-block edge. Let the within-block degree distribution be $p_w(k)$, the across block degree distribution be $p_a(k)$, $p'_w(k) = \frac{k}{c_w}p_w(k)$ be the within-block degree distribution of a node reached via a within-block edge, and $p'_a(k) = \frac{k}{c_a}p_a(k)$ be the across-block degree distribution of a node reached via an across-block edge, with $c_w$ and $c_a$ being normalizing constants.

We first determine the probability a node is reached via a within-block edge. Let the event a node is reached via a within-block edge be $W$, and let the event the node's parent is reached via a within-block edge and across-block edge be $W_P$ and $A_P$, respectively. Then:
\begin{align*}
    P(W) = P(W | W_P) P(W_P) + P(W | A_P) P(A_P),
\end{align*}
with the observation that $P(W_P) = P(W)$ and $P(A_P) = 1-P(W)$. To determine the conditional probabilities, we have
\begin{align*}
    P(W | W_P) &= 
    \sum_{0 < k_w < N-1} \sum_{0 < k_a < N-1} \frac{k_wp_w(k_w)}{c_w} p_a(k_a) \frac{k_w - 1}{k_w + k_a - 1}, \\
    P(W | A_P) &= 
    \sum_{0 < k_w < N-1} \sum_{0 < k_a < N-1} p_w(k_w) \frac{k_ap_a(k_a)}{c_a} \frac{k_w}{k_w + k_a - 1},
\end{align*}
from the law of total probability: for each possible within-block and across-block degree of the parent node, we find the probability that the parent had that probability (product of the first two terms, with the degree distribution of the reaching type of edge scaled by degree and normalized) and the probability that a within-block edge is traversed (last term). With this we may solve for $P(W)$.

We now build degree distributions for the ``average'' node in the sample. For example, the within-block degree distribution is a mixture of $p_w(k)$ with weight $1-P(W)$ and $p_w'(k)$ with weight $P(W)$: with probability $P(W)$ the node is reached via a within-block edge and so it's within-block degree comes from the weighted distribution, and with probability $1-P(W)$ the node is reached via an across-block edge and it's within-block degree comes from the unweighted distribution. We again express this average distribution using polynomials. Let
\begin{align*}
    q_w(x) &= \sum_{k=0}^\infty p_w(k)x^k, q_a(x) = \sum_{k=0}^\infty p_a(k)x^k, \\
    q'_w(x) &= \sum_{k=0}^\infty p'_w(k)x^k, q'_a(x) = \sum_{k=0}^\infty p'_a(k)x^k.
\end{align*}
Then 
\begin{align*}
    q^*_w(x) &= (1-P(W)) q_w(x) + P(W) q'_w(x)/x^2 \\
    q^*_a(x) &= P(W) q_a(x) + (1-P(W)) q'_a(x)/x^2
\end{align*}
are the polynomials representing the average distribution within the sample. Here we apply the correction with the $/x^2$ term; we reduce the degree contribution to the corresponding edge type by $2$ when the edge type is fully inside the sample. We may now conclude as in the random sampling case: $q^*_{Sw}(x) = q^*_w(x)^s$ with $p_{Sw}(k)$ approximately the coefficient on $x^k$, and similar for $p_{Wa}(k)$ and $q^*_{Sa}(x) = q^*_{Sa}(x)^s$.

\subsection{Experimental Design}

We run extensive experiments on these settings. For each graph setting, we generate $5$ graphs. On these graphs, for each sampling method and sample size, we generate $20$ samples and calculate the resulting DSPD. For each graph setting, sampling method, and sample size, we collect timing statistics on the $100$ runs and average the derived DSPDs to obtain our empirical distribution. We also run our framework $100$ times and collect timing information; however each run of our framework yields the same estimated DSPD.

\subsubsection{Downstream Task}

Our downstream task investigates which sampling method gives a smaller mean shortest-path distance to the sample for a given graph setting and sample size. This metric is of interest because a smaller mean distance can indicate better average performance for graph neural networks~\cite{ma2021subgroup} and the first moment of a distribution is an important measure in general.

Recall that the Power Law A and Power Law B configurations were chosen to provide different preferences on this downstream task. In the Power Law A configurations snowball sampling consistently gives smaller mean distances than random sampling, and vice versa in the Power Law B configurations.

\section{Results}
\label{sec:results}

\begin{table*}
\caption{Wasserstein-1 distances of framework estimation to empirically obtained distributions. Means and standard deviations over 100 trials reported. Settings described as \texttt{[graph type] [graph size] [sampling method] [sample size]}. \texttt{bin} is binomial, \texttt{pow} is power, \texttt{rand} is random, \texttt{snow} is snowball.}
\label{tab:distances}
\begin{tabular}{lc|lc}
\toprule
Setting & Wasserstein-1 Distance & Setting & Wasserstein-1 Distance\\
\midrule
\texttt{bin 20000 rand 200} & $0.0153 \pm 0.0089$ & \texttt{pow\_b 20000 rand 200} & $0.0246 \pm 0.0104$ \\
\texttt{bin 20000 rand 400} & $0.0227 \pm 0.0057$ & \texttt{pow\_b 20000 rand 400} & $0.0319 \pm 0.0080$ \\
\texttt{bin 20000 rand 1000} & $0.0120 \pm 0.0022$ & \texttt{pow\_b 20000 rand 1000} & $0.0170 \pm 0.0044$ \\
\texttt{bin 20000 snow 200} & $0.0120 \pm 0.0069$ & \texttt{pow\_b 20000 snow 200} & $0.0251 \pm 0.0118$ \\
\texttt{bin 20000 snow 400} & $0.0181 \pm 0.0056$ & \texttt{pow\_b 20000 snow 400} & $0.0328 \pm 0.0085$ \\
\texttt{bin 20000 snow 1000} & $0.0151 \pm 0.0025$ & \texttt{pow\_b 20000 snow 1000} & $0.0230 \pm 0.0056$ \\
\texttt{bin 40000 rand 200} & $0.0174 \pm 0.0051$ & \texttt{pow\_b 40000 rand 200} & $0.0353 \pm 0.0110$ \\
\texttt{bin 40000 rand 400} & $0.0139 \pm 0.0078$ & \texttt{pow\_b 40000 rand 400} & $0.0273 \pm 0.0097$ \\
\texttt{bin 40000 rand 1000} & $0.0231 \pm 0.0042$ & \texttt{pow\_b 40000 rand 1000} & $0.0339 \pm 0.0057$ \\
\texttt{bin 40000 snow 200} & $0.0198 \pm 0.0045$ & \texttt{pow\_b 40000 snow 200} & $0.0372 \pm 0.0124$ \\
\texttt{bin 40000 snow 400} & $0.0112 \pm 0.0064$ & \texttt{pow\_b 40000 snow 400} & $0.0297 \pm 0.0107$ \\
\texttt{bin 40000 snow 1000} & $0.0207 \pm 0.0040$ & \texttt{pow\_b 40000 snow 1000} & $0.0397 \pm 0.0067$ \\
\texttt{bin 100000 rand 200} & $0.0248 \pm 0.0083$ & \texttt{pow\_b 100000 rand 200} & $0.0407 \pm 0.0126$ \\
\texttt{bin 100000 rand 400} & $0.0224 \pm 0.0055$ & \texttt{pow\_b 100000 rand 400} & $0.0419 \pm 0.0074$ \\
\texttt{bin 100000 rand 1000} & $0.0195 \pm 0.0043$ & \texttt{pow\_b 100000 rand 1000} & $0.0315 \pm 0.0051$ \\
\texttt{bin 100000 snow 200} & $0.0179 \pm 0.0059$ & \texttt{pow\_b 100000 snow 200} & $0.0420 \pm 0.0159$ \\
\texttt{bin 100000 snow 400} & $0.0217 \pm 0.0028$ & \texttt{pow\_b 100000 snow 400} & $0.0474 \pm 0.0114$ \\
\texttt{bin 100000 snow 1000} & $0.0125 \pm 0.0048$ & \texttt{pow\_b 100000 snow 1000} & $0.0354 \pm 0.0071$ \\
\midrule
\texttt{pow\_a 20000 rand 200} & $0.0612 \pm 0.0137$ & \texttt{sbm 20000 rand 200} & $0.1499 \pm 0.0175$ \\
\texttt{pow\_a 20000 rand 400} & $0.0541 \pm 0.0101$ & \texttt{sbm 20000 rand 400} & $0.0581 \pm 0.0095$ \\
\texttt{pow\_a 20000 rand 1000} & $0.0242 \pm 0.0057$ & \texttt{sbm 20000 rand 1000} & $0.0050 \pm 0.0026$ \\
\texttt{pow\_a 20000 snow 200} & $0.0651 \pm 0.0163$ & \texttt{sbm 20000 snow 200} & $0.7078 \pm 0.0624$ \\
\texttt{pow\_a 20000 snow 400} & $0.0536 \pm 0.0090$ & \texttt{sbm 20000 snow 400} & $0.5668 \pm 0.0404$ \\
\texttt{pow\_a 20000 snow 1000} & $0.0318 \pm 0.0057$ & \texttt{sbm 20000 snow 1000} & $0.3768 \pm 0.0298$ \\
\texttt{pow\_a 40000 rand 200} & $0.0475 \pm 0.0153$ & \texttt{sbm 40000 rand 200} & $0.1469 \pm 0.0116$ \\
\texttt{pow\_a 40000 rand 400} & $0.0686 \pm 0.0106$ & \texttt{sbm 40000 rand 400} & $0.1074 \pm 0.0100$ \\
\texttt{pow\_a 40000 rand 1000} & $0.0478 \pm 0.0061$ & \texttt{sbm 40000 rand 1000} & $0.0188 \pm 0.0037$ \\
\texttt{pow\_a 40000 snow 200} & $0.0513 \pm 0.0172$ & \texttt{sbm 40000 snow 200} & $0.4777 \pm 0.0395$ \\
\texttt{pow\_a 40000 snow 400} & $0.0742 \pm 0.0109$ & \texttt{sbm 40000 snow 400} & $0.4406 \pm 0.0265$ \\
\texttt{pow\_a 40000 snow 1000} & $0.0493 \pm 0.0063$ & \texttt{sbm 40000 snow 1000} & $0.2844 \pm 0.0181$ \\
\texttt{pow\_a 100000 rand 200} & $0.0566 \pm 0.0135$ & \texttt{sbm 100000 rand 200} & $0.0162 \pm 0.0051$ \\
\texttt{pow\_a 100000 rand 400} & $0.0469 \pm 0.0084$ & \texttt{sbm 100000 rand 400} & $0.0196 \pm 0.0049$ \\
\texttt{pow\_a 100000 rand 1000} & $0.0742 \pm 0.0064$ & \texttt{sbm 100000 rand 1000} & $0.0220 \pm 0.0023$ \\
\texttt{pow\_a 100000 snow 200} & $0.0519 \pm 0.0119$ & \texttt{sbm 100000 snow 200} & $0.1518 \pm 0.0146$ \\
\texttt{pow\_a 100000 snow 400} & $0.0521 \pm 0.0107$ & \texttt{sbm 100000 snow 400} & $0.1666 \pm 0.0179$ \\
\texttt{pow\_a 100000 snow 1000} & $0.0822 \pm 0.0080$ & \texttt{sbm 100000 snow 1000} & $0.1331 \pm 0.0112$ \\
\bottomrule
\end{tabular}
\end{table*}

\begin{figure}[hbtp]
    \centering
    \begin{subfigure}[t]{.45\linewidth}
        \centering
        \includegraphics[width=1.\linewidth]{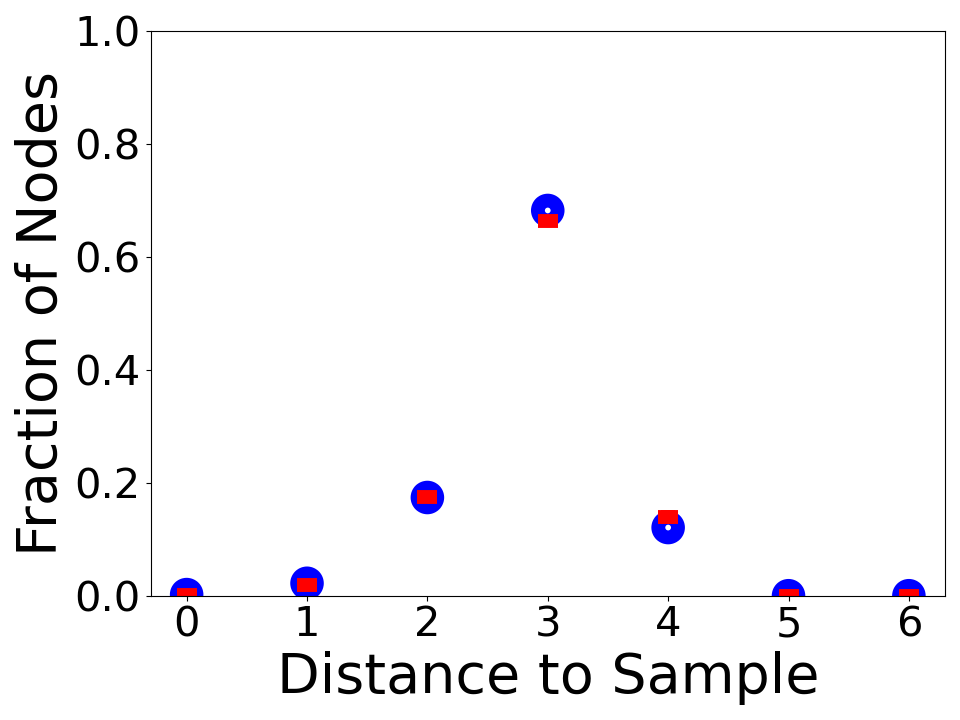}
        \caption{\texttt{bin 100000 rand 200}\\($d=0.0248$)}
        \label{fig:accuracy_bin}
    \end{subfigure}
    ~
    \begin{subfigure}[t]{.45\linewidth}
        \centering
        \includegraphics[width=1.\linewidth]{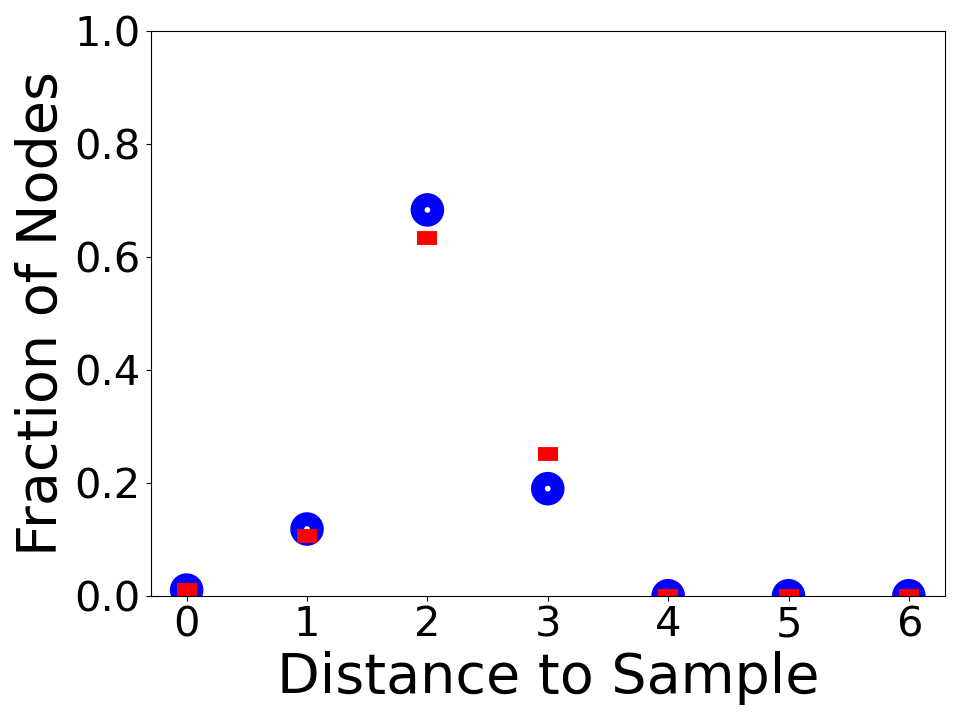}
        \caption{\texttt{pow\_a 100000 snow 1000}\\($d=0.0822$)}
        \label{fig:accuracy_pow_a}
    \end{subfigure}
    \\
    \begin{subfigure}[t]{.45\linewidth}
        \centering
        \includegraphics[width=1.\linewidth]{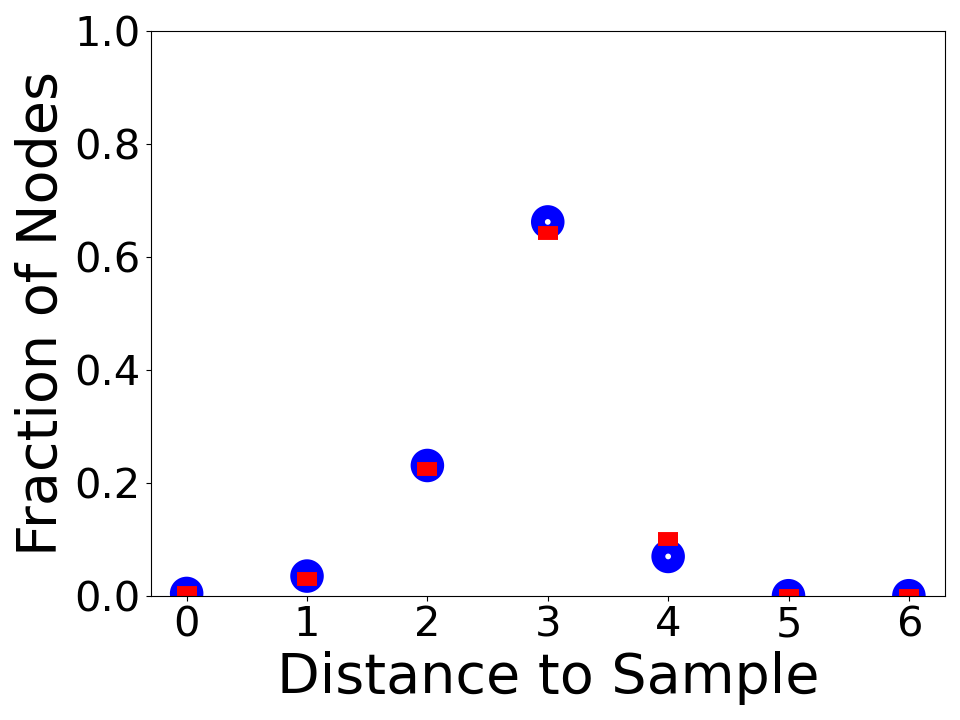}
        \caption{\texttt{pow\_b 100000 snow 400}\\($d=0.0474)$}
        \label{fig:accuracy_pow_b}
    \end{subfigure}
    ~
    \begin{subfigure}[t]{.45\linewidth}
        \centering
        \includegraphics[width=1.\linewidth]{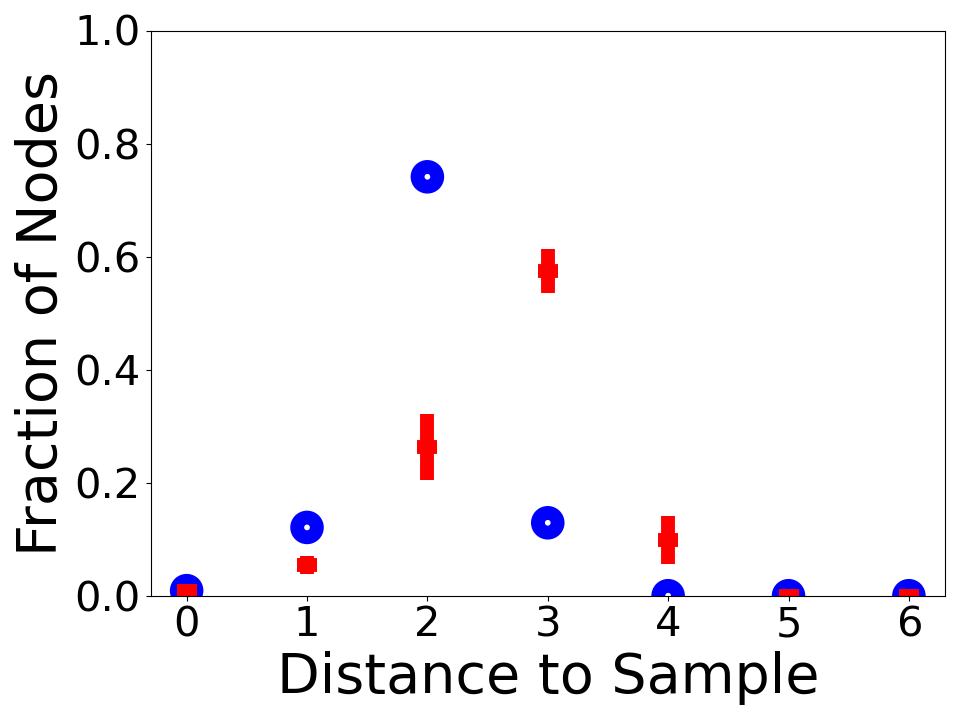}
        \caption{\texttt{sbm 20000 snow 200}\\($d=0.7078$)}
        \label{fig:accuracy_sbm_1}
    \end{subfigure}
    \\
    \begin{subfigure}[t]{.45\linewidth}
        \centering
        \includegraphics[width=1.\linewidth]{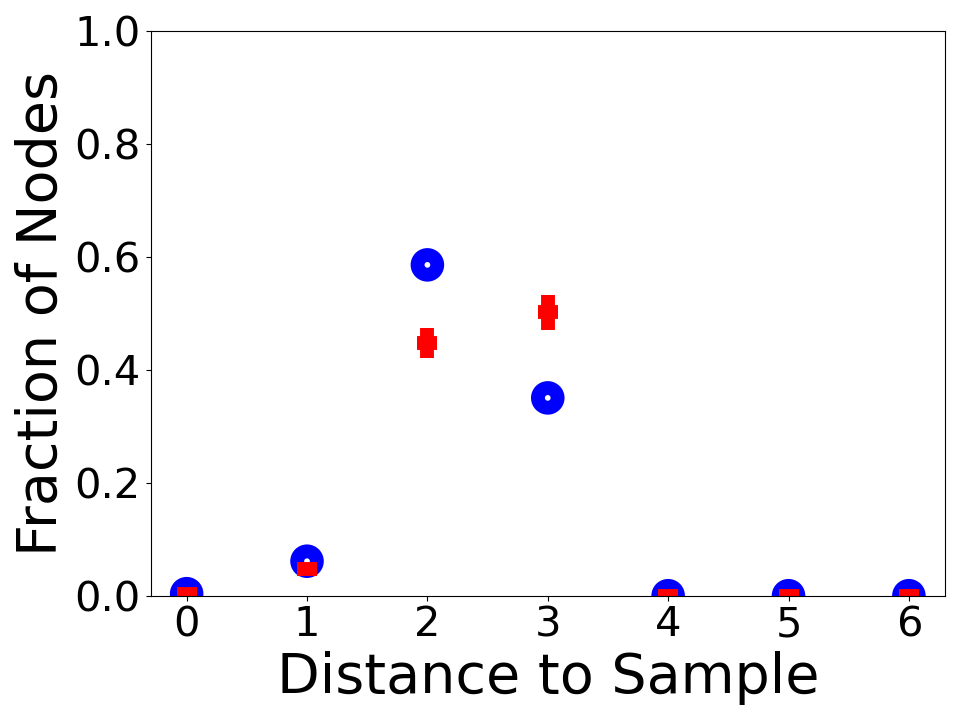}
        \caption{\texttt{sbm 100000 snow 400}\\($d=0.1666$)}
        \label{fig:accuracy_sbm_2}
    \end{subfigure}
    ~
    \begin{subfigure}[t]{.45\linewidth}
        \centering
        \includegraphics[width=1.\linewidth]{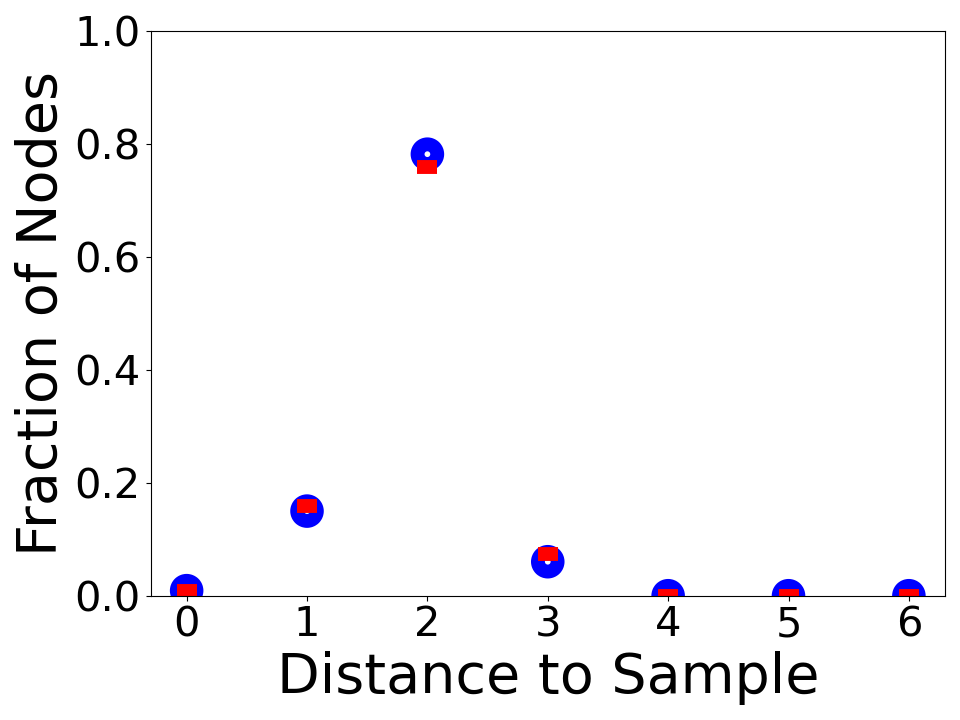}
        \caption{\texttt{sbm 100000 rand 1000}\\($d=0.0220$)}
        \label{fig:accuracy_sbm_3}
    \end{subfigure}
    \caption{Visual representations of accuracies of estimated DSPDs. Blue circles mark the estimated DSPD; red lines mark the empirical DSPD, with $2\times$ standard errors marked vertically. Selections for \texttt{bin}, \texttt{pow\_a}, and \texttt{pow\_b} are for least accurate estimates in each graph types; selections for \texttt{sbm} are for least accurate estimate overall and for random and snowball sampling on graph size $100000$.}
    \Description{Visual representations of accuracies of estimated distributions of shortest-path distances. Estimations are highly accurate when compared to empirical distributions in binomial and power law graphs, as well as large stochastic block model graphs under random sampling. Estimation is inaccurate for small stochastic block model graphs under snowball sampling, but accuracy improves for large graph sizes.}
    \label{fig:accuracy}
\end{figure}

\begin{figure}[hbtp]
    \centering
    \includegraphics[width=\linewidth]{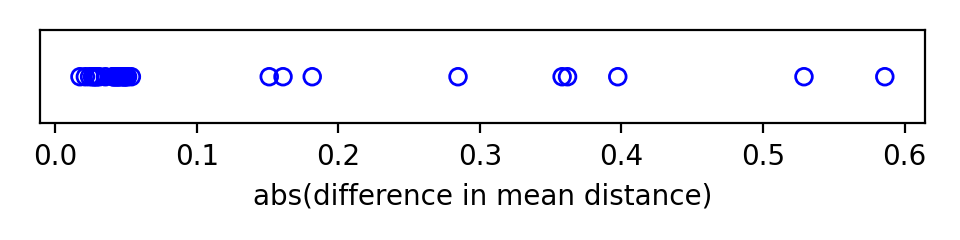}
    \caption{Absolute difference of mean distances between random and snowball sampling for evaluated configurations.}
    \Description{Scatterplot of absolute differences in mean distance. Most differences are less than 0.1, and the minimum is very close to zero. The maximum difference extends up to about 0.6.}
    \label{fig:diff_mean}
\end{figure}

\begin{table*}
\caption{Time to obtain distributions through empirical methods and estimation via our framework. Means and standard deviations over 100 trials reported. Faster times bolded. Settings described as \texttt{[graph type] [graph size] [sampling method] [sample size]}. \texttt{bin} is binomial, \texttt{pow} is power, \texttt{rand} is random, \texttt{snow} is snowball.}
\label{tab:timings}
\begin{tabular}{lcc|lcc}
\toprule
Description & Empirical & Framework & Description & Empirical & Framework \\
\midrule
\texttt{bin 20000 rand 200} & $0.0443 \pm 0.0034$ & $\mathbf{0.0080 \pm 0.0018}$ & \texttt{pow\_b 20000 rand 200} & $0.0476 \pm 0.0036$ & $\mathbf{0.0071 \pm 0.0014}$ \\
\texttt{bin 20000 rand 400} & $0.0551 \pm 0.0038$ & $\mathbf{0.0320 \pm 0.0070}$ & \texttt{pow\_b 20000 rand 400} & $0.0501 \pm 0.0037$ & $\mathbf{0.0254 \pm 0.0032}$ \\
\texttt{bin 20000 rand 1000} & $\mathbf{0.0612 \pm 0.0113}$ & $0.1622 \pm 0.0083$ & \texttt{pow\_b 20000 rand 1000} & $\mathbf{0.0526 \pm 0.0041}$ & $0.1438 \pm 0.0026$ \\
\texttt{bin 20000 snow 200} & $0.0531 \pm 0.0033$ & $\mathbf{0.0112 \pm 0.0028}$ & \texttt{pow\_b 20000 snow 200} & $0.0490 \pm 0.0032$ & $\mathbf{0.0082 \pm 0.0020}$ \\
\texttt{bin 20000 snow 400} & $0.0587 \pm 0.0112$ & $\mathbf{0.0278 \pm 0.0019}$ & \texttt{pow\_b 20000 snow 400} & $0.0496 \pm 0.0037$ & $\mathbf{0.0246 \pm 0.0008}$ \\
\texttt{bin 20000 snow 1000} & $\mathbf{0.0530 \pm 0.0033}$ & $0.1900 \pm 0.0172$ & \texttt{pow\_b 20000 snow 1000} & $\mathbf{0.0501 \pm 0.0037}$ & $0.1471 \pm 0.0093$ \\
\texttt{bin 40000 rand 200} & $0.0954 \pm 0.0064$ & $\mathbf{0.0088 \pm 0.0015}$ & \texttt{pow\_b 40000 rand 200} & $0.1062 \pm 0.0060$ & $\mathbf{0.0067 \pm 0.0002}$ \\
\texttt{bin 40000 rand 400} & $0.1168 \pm 0.0061$ & $\mathbf{0.0275 \pm 0.0009}$ & \texttt{pow\_b 40000 rand 400} & $0.1133 \pm 0.0063$ & $\mathbf{0.0253 \pm 0.0022}$ \\
\texttt{bin 40000 rand 1000} & $\mathbf{0.1183 \pm 0.0065}$ & $0.1581 \pm 0.0065$ & \texttt{pow\_b 40000 rand 1000} & $\mathbf{0.1142 \pm 0.0058}$ & $0.1453 \pm 0.0052$ \\
\texttt{bin 40000 snow 200} & $0.1170 \pm 0.0052$ & $\mathbf{0.0076 \pm 0.0005}$ & \texttt{pow\_b 40000 snow 200} & $0.1118 \pm 0.0060$ & $\mathbf{0.0077 \pm 0.0003}$ \\
\texttt{bin 40000 snow 400} & $0.1183 \pm 0.0076$ & $\mathbf{0.0278 \pm 0.0027}$ & \texttt{pow\_b 40000 snow 400} & $0.1107 \pm 0.0054$ & $\mathbf{0.0249 \pm 0.0025}$ \\
\texttt{bin 40000 snow 1000} & $\mathbf{0.1170 \pm 0.0060}$ & $0.1570 \pm 0.0049$ & \texttt{pow\_b 40000 snow 1000} & $\mathbf{0.1118 \pm 0.0057}$ & $0.1442 \pm 0.0040$ \\
\texttt{bin 100000 rand 200} & $0.2680 \pm 0.0102$ & $\mathbf{0.0082 \pm 0.0020}$ & \texttt{pow\_b 100000 rand 200} & $0.3023 \pm 0.0108$ & $\mathbf{0.0068 \pm 0.0004}$ \\
\texttt{bin 100000 rand 400} & $0.3269 \pm 0.0120$ & $\mathbf{0.0310 \pm 0.0007}$ & \texttt{pow\_b 100000 rand 400} & $0.3152 \pm 0.0179$ & $\mathbf{0.0263 \pm 0.0015}$ \\
\texttt{bin 100000 rand 1000} & $0.3263 \pm 0.0097$ & $\mathbf{0.1707 \pm 0.0046}$ & \texttt{pow\_b 100000 rand 1000} & $0.2984 \pm 0.0122$ & $\mathbf{0.1432 \pm 0.0039}$ \\
\texttt{bin 100000 snow 200} & $0.3254 \pm 0.0119$ & $\mathbf{0.0081 \pm 0.0012}$ & \texttt{pow\_b 100000 snow 200} & $0.3028 \pm 0.0091$ & $\mathbf{0.0067 \pm 0.0004}$ \\
\texttt{bin 100000 snow 400} & $0.3227 \pm 0.0103$ & $\mathbf{0.0273 \pm 0.0004}$ & \texttt{pow\_b 100000 snow 400} & $0.3238 \pm 0.0429$ & $\mathbf{0.0261 \pm 0.0019}$ \\
\texttt{bin 100000 snow 1000} & $0.3257 \pm 0.0110$ & $\mathbf{0.1554 \pm 0.0050}$ & \texttt{pow\_b 100000 snow 1000} & $0.2960 \pm 0.0148$ & $\mathbf{0.1437 \pm 0.0043}$ \\
\midrule
\texttt{pow\_a 20000 rand 200} & $0.0523 \pm 0.0042$ & $\mathbf{0.0109 \pm 0.0005}$ & \texttt{sbm 20000 rand 200} & $0.0518 \pm 0.0046$ & $\mathbf{0.0137 \pm 0.0022}$ \\
\texttt{pow\_a 20000 rand 400} & $0.0559 \pm 0.0028$ & $\mathbf{0.0433 \pm 0.0033}$ & \texttt{sbm 20000 rand 400} & $0.0617 \pm 0.0042$ & $\mathbf{0.0395 \pm 0.0014}$ \\
\texttt{pow\_a 20000 rand 1000} & $\mathbf{0.0571 \pm 0.0037}$ & $0.2647 \pm 0.0078$ & \texttt{sbm 20000 rand 1000} & $\mathbf{0.0641 \pm 0.0051}$ & $0.2217 \pm 0.0093$ \\
\texttt{pow\_a 20000 snow 200} & $0.0559 \pm 0.0128$ & $\mathbf{0.0107 \pm 0.0003}$ & \texttt{sbm 20000 snow 200} & $0.0603 \pm 0.0028$ & $\mathbf{0.0139 \pm 0.0021}$ \\
\texttt{pow\_a 20000 snow 400} & $0.0555 \pm 0.0120$ & $\mathbf{0.0423 \pm 0.0031}$ & \texttt{sbm 20000 snow 400} & $0.0605 \pm 0.0038$ & $\mathbf{0.0468 \pm 0.0005}$ \\
\texttt{pow\_a 20000 snow 1000} & $\mathbf{0.0556 \pm 0.0121}$ & $0.2643 \pm 0.0082$ & \texttt{sbm 20000 snow 1000} & $\mathbf{0.0613 \pm 0.0051}$ & $0.2670 \pm 0.0093$ \\
\texttt{pow\_a 40000 rand 200} & $0.1144 \pm 0.0050$ & $\mathbf{0.0115 \pm 0.0015}$ & \texttt{sbm 40000 rand 200} & $0.1101 \pm 0.0048$ & $\mathbf{0.0114 \pm 0.0008}$ \\
\texttt{pow\_a 40000 rand 400} & $0.1227 \pm 0.0054$ & $\mathbf{0.0435 \pm 0.0034}$ & \texttt{sbm 40000 rand 400} & $0.1346 \pm 0.0073$ & $\mathbf{0.0409 \pm 0.0057}$ \\
\texttt{pow\_a 40000 rand 1000} & $\mathbf{0.1230 \pm 0.0058}$ & $0.2620 \pm 0.0063$ & \texttt{sbm 40000 rand 1000} & $\mathbf{0.1371 \pm 0.0053}$ & $0.2255 \pm 0.0137$ \\
\texttt{pow\_a 40000 snow 200} & $0.1216 \pm 0.0059$ & $\mathbf{0.0107 \pm 0.0004}$ & \texttt{sbm 40000 snow 200} & $0.1348 \pm 0.0062$ & $\mathbf{0.0135 \pm 0.0009}$ \\
\texttt{pow\_a 40000 snow 400} & $0.1206 \pm 0.0056$ & $\mathbf{0.0426 \pm 0.0035}$ & \texttt{sbm 40000 snow 400} & $0.1345 \pm 0.0042$ & $\mathbf{0.0469 \pm 0.0014}$ \\
\texttt{pow\_a 40000 snow 1000} & $\mathbf{0.1214 \pm 0.0063}$ & $0.2620 \pm 0.0069$ & \texttt{sbm 40000 snow 1000} & $\mathbf{0.1342 \pm 0.0056}$ & $0.2671 \pm 0.0103$ \\
\texttt{pow\_a 100000 rand 200} & $0.3166 \pm 0.0110$ & $\mathbf{0.0116 \pm 0.0019}$ & \texttt{sbm 100000 rand 200} & $0.2936 \pm 0.0114$ & $\mathbf{0.0110 \pm 0.0005}$ \\
\texttt{pow\_a 100000 rand 400} & $0.3364 \pm 0.0135$ & $\mathbf{0.0436 \pm 0.0007}$ & \texttt{sbm 100000 rand 400} & $0.4294 \pm 0.0503$ & $\mathbf{0.0399 \pm 0.0035}$ \\
\texttt{pow\_a 100000 rand 1000} & $0.3387 \pm 0.0109$ & $\mathbf{0.2605 \pm 0.0056}$ & \texttt{sbm 100000 rand 1000} & $0.4128 \pm 0.0111$ & $\mathbf{0.2195 \pm 0.0102}$ \\
\texttt{pow\_a 100000 snow 200} & $0.3347 \pm 0.0118$ & $\mathbf{0.0112 \pm 0.0005}$ & \texttt{sbm 100000 snow 200} & $0.4174 \pm 0.0205$ & $\mathbf{0.0133 \pm 0.0004}$ \\
\texttt{pow\_a 100000 snow 400} & $0.3353 \pm 0.0099$ & $\mathbf{0.0423 \pm 0.0025}$ & \texttt{sbm 100000 snow 400} & $0.4168 \pm 0.0139$ & $\mathbf{0.0472 \pm 0.0030}$ \\
\texttt{pow\_a 100000 snow 1000} & $0.3331 \pm 0.0105$ & $\mathbf{0.2599 \pm 0.0062}$ & \texttt{sbm 100000 snow 1000} & $0.4179 \pm 0.0143$ & $\mathbf{0.2643 \pm 0.0064}$ \\
\bottomrule
\end{tabular}
\end{table*}

\subsection{Accuracy of Framework}

We evaluate the accuracy of our framework by examining the Wasserstein-1 distance between the distribution estimated by our framework and actual distributions obtained through sampling. Intuitively, the Wasserstein-1 distance measures the ``amount'' of probability that needs to be moved to make two distributions identical, weighted by the ``distance'' it needs to be moved. We report the mean distance and the standard deviation of the distances for all settings in Table \ref{tab:distances}. We also provide plots visually illustrating the differences in distributions for different Wasserstein-1 distances in Figure \ref{fig:accuracy}.

Our framework demonstrates high accuracy in the binomial and power-law settings, with mean Wasserstein-1 distances less than $0.09$ in all cases. Our framework is especially accurate for binomial graphs, where the mean Wasserstein-1 distance is always less than $0.03$. There does not appear to be any differences in accuracy with respect to graph size, sampling method, or sampling size. Figures \ref{fig:accuracy_bin}, \ref{fig:accuracy_pow_a}, and \ref{fig:accuracy_pow_b} visually illustrate the accuracy associated with these Wasserstein-1 distances.

For SBM graphs, the framework's accuracy decreases under certain conditions: for a graph size of $20000$ and a snowball sample of size $200$, the mean Wasserstein-1 distance is $0.7078$ (see Figure \ref{fig:accuracy_sbm_1}). In general on SBM graphs, the framework is less accurate when the graph is smaller, when the sample size is smaller and/or when using snowball sampling. However, the accuracy does improve with increasing graph size: at size $100000$, the Wasserstein-1 distance for random sampling is less than $0.03$ compared to $0.2$ for smaller graph sizes, and the Wasserstein-1 distance for snowball sampling is less than $0.2$ compared to $0.8$ for smaller graph sizes. Figures \ref{fig:accuracy_sbm_2} and \ref{fig:accuracy_sbm_3} visually illustrate these accuracies at larger graph sizes.

These results suggest that the presence of community structures poses a challenge for the framework. In binomial and power law graphs, there is no community structure, and our framework is very accurate. In SBM graphs, there is community structure, and this is exasperated under snowball sampling, where nodes in the sample are likely to come from the same community. This may be why our framework is more accurate under random sampling than snowball sampling in SBM graphs: random sampling ameliorates some of the difficulty from community structures. Interestingly, our framework appears to be more accurate as graph size increases, indicating that it will remain useful in the domains where graph sampling is of most use: very large graph sizes. Still, we show that despite poor accuracy in some cases, there is still useful information to be gleaned from the estimates.

\subsubsection{Performance on Downstream Task}

The framework proves to be a reliable estimator for DSPD in binomial and power-law graphs, and certain configurations and sampling methods on SBM graphs. However, even on the SBM graph configurations where our framework does not produce an accurate estimate, the estimate is still effective at downstream tasks. Specifically, we evaluate how accurately our model can compare the mean distances resulting from random or snowball sampling for a fixed graph configuration and sample size. This is analogous to using our framework to choose a sampling method for a specific network and sampling budget.

Our framework accurately compares mean distances across all graph configurations and sampling methods, including for the SBM graphs where the immediate estimation is inaccurate. For all graph configurations and sample sizes, the mean distance comparison in the estimated distributions matched exactly with the mean distance comparison in the empirical distributions. These include cases where snowball sampling gives smaller mean distance (e.g., Power Law A) and cases where random sampling gives smaller mean distance (e.g., Power Law B). Additionally, this method is very sensitive, able to correctly compare distributions when most differences are less than $0.1$. The minimum difference was $0.0175$. A scatterplot of all the differences is presented in Figure \ref{fig:diff_mean}.

\subsection{Efficiency of Framework}

The framework demonstrates significantly greater efficiency compared to empirically determining the DSPD to a sample, which requires first obtaining a graph then performing sampling, at a minimum. This can be costly even in simulation settings, and more so in the real world. On the other hand, our framework only requires specifying a graph configuration, sampling method, and sample size, with no need to obtain or produce a graph or sample. For example, it took around 8 minutes to generate a binomial graph of size $100000$ to obtain empirical distributions, but this step is not necessary for our framework. However, even when just considering the cost to calculate the distribution (i.e., timing after having the graph and sample), our framework has good efficiency.

Table \ref{tab:timings} describes the time taken to calculate a single DSPD, with means and standard deviations over $100$ trials. In practice, to empirically determine the DSPD one would want to take an average over multiple trials to account for randomness in graph generation and sample selection, but here we just report the timings for a single trial. In nearly all scenarios, the framework consistently outperforms empirical methods in terms of speed, except for small graphs with large sample sizes. In some cases, our framework can be an order of magnitude faster (e.g., Power A graphs of size $40000$ with a snowball sample of size $200$). 

These results align with expectations given algorithm implementations. Empirical distributions can be determined via a breadth-first search, which scales linearly with graph size. Our framework employs polynomial exponentiation with order equal to the sample size, which scales quadratically with sample size but is constant with graph size. 
Thus, our framework is most performant for very large graphs with small sample sizes, which is a common use case for graph sampling. 

\section{Conclusion}
\label{sec:conclusion}

We propose a framework to accurately and efficiently estimate the DSPD to a sample in a graph by extending an existing framework that estimates the DSPD to a single random node. This framework does not require access to the full graph structure or the sample; rather, it only requires a model of the graph and the sampling method to obtain relevant degree distributions.

We evaluate the accuracy and efficiency of our framework by applying it to graphs with and without community structures and to random and snowball sampling. Our evaluation shows that the proposed framework is generally accurate, achieving high accuracy compared to empirically obtained distributions on graphs without community structures and large graphs with community structures. Even when accuracy decreases due to the complexity of handling community structures in smaller graphs, the framework remains useful for downstream tasks, consistently achieving perfect comparisons of mean distances across different sampling methods. Moreover, our framework is highly efficient. It eliminates the need to generate graphs and samples, and even when empirical methods have access to them, our framework can be up to an order of magnitude faster. While empirical methods scale linearly with graph size, our framework scales quadratically with sample size, making it especially effective for large graphs and small sample sizes. These advantages suggest that the proposed framework could be used to evaluate and select sampling methods for very large graphs without needing to perform the sampling itself. Future work will validate this approach using real-world graph datasets.

\bibliographystyle{ACM-Reference-Format}
\bibliography{shortest_paths}

\end{document}